\documentclass[12pt]{iopart}
\usepackage{graphicx}
\usepackage{times}

\begin{document}

\title[The Gaussian formula and spherical aberrations from Fermat's principle]{The Gaussian formula and spherical aberrations of the static and relativistic curved mirrors from Fermat's principle}
\author{Sylvia H.\ Sutanto$^1$ and Paulus C.\ Tjiang$^2$}
\address{Theoretical and Computational Physics Group, Department of Physics, Faculty of Information Technology and Sciences, Parahyangan Catholic University, Bandung 40141 - INDONESIA}
\ead{$^1$sylvia@home.unpar.ac.id, $^2$pctjiang@home.unpar.ac.id}

\begin{abstract}
The Gaussian formula and spherical aberrations of the static and relativistic curved mirrors are analyzed using the optical path length (OPL) and Fermat's principle. The geometrical figures generated by the rotation of conic sections about their symmetry axes are considered for the shapes of the mirrors. By comparing the results in static and relativistic cases, it is shown that the focal lengths and the spherical aberration relations of the relativistic mirrors obey the Lorentz contraction. Further analysis of the spherical aberrations for both static and relativistic cases have resulted in the information about the limits for the paraxial approximation, as well as for the minimum speed of the systems to reduce the spherical aberrations.
\end{abstract}


\section{Introduction}
\label{Intro}

In every discussion of geometrical optics, the so-called Fermat's principle hardly misses its part. Many textbooks on either elementary physics or optics discuss the application of Fermat's principle to derive the laws of reflection and refraction~\cite{physicsbooks}, but only few of them discuss the application of the principle on curved mirrors and lenses. To the best of our knowledge, the first discussion of the application of Fermat's principle on curved mirrors and lenses, particularly those of the spherical shapes, was given by Lemons with his proposal of the time delay function~\cite{Lemons94}. A more general treatment of the time delay function was given by Lakshminarayanan et. al.~\cite{Lakshmi01}, and a non-technical discussion of curved mirrors and Fermat's principle was presented by Erb~\cite{Erb95}.

On the other hand, discussion of the relativistic plane mirrors was recently brought up by Gjurchinovski, both using Huygens construction~\cite{Gjurchinovski04} and Fermat's principle~\cite{Gjurchinovski04-2}. The same discussion was presented for the first time more than 100 years ago by Einstein through his famous work on the special theory of relativity~\cite{Einstein05}. However, to the best of our knowledge, no similar discussion on relativistic curved mirrors exists.

The purpose of this work is to give an analysis of relativistic curved mirrors and compare the results to static case. Although spherical and parabolic shapes of mirrors are commonly used in practical applications, we shall consider the shapes of curved mirrors generated by the rotation of the  two dimensional conic sections about their symmetry axes, inspired by Watson~\cite{Watson99}. The straightforward calculations are performed (without the introduction of a time delay function) to provide clearer interpretations on the results. The results obtained in this work might be used to make relativistic corrections to the images of light emitting astronomical objects moving at very high speeds, which are formed by some optical apparati, especially those that use curved mirrors, such as reflector telescopes.

The paper is organized as follows : Section~\ref{conic-section} reviews the general mathematical form of the conic sections, which will be used to shape the curved mirrors. In this case, we shall follow the discussions of conic sections provided by Watson~\cite{Watson99} and Baker~\cite{Baker43}. In Section~\ref{static-mirror} we shall discuss the application of Fermat's principle to obtain Gaussian formula and spherical aberration relations of static curved mirrors, with some analysis of the focal length of the mirror and the physical behaviour of the spherical aberrations. An analogous discussion for relativistic curved mirrors will be given in Section~\ref{moving-mirror}, with some analysis of the results compared to static case. The discussion concludes in Section~\ref{conclusion}, with some remarks on future work for relativistic curved lenses.

\section{General curved mirror : the conic sections}
\label{conic-section}

Let us consider curved mirrors whose shapes are the geometrical figures generated from the rotation of two dimensional conic sections about their symmetry axes. These shapes are well known in the construction of astronomical instruments~\cite{Watson99}. It is worth noting that the discussion of conic sections is useful for construction of mathematical models to predict the theoretical change of corneal asphericity after the eye surgery~\cite{Gantinel01}, although it is beyond our scope of discussion.

\begin{figure}
\begin{center}
\includegraphics[scale=0.6]{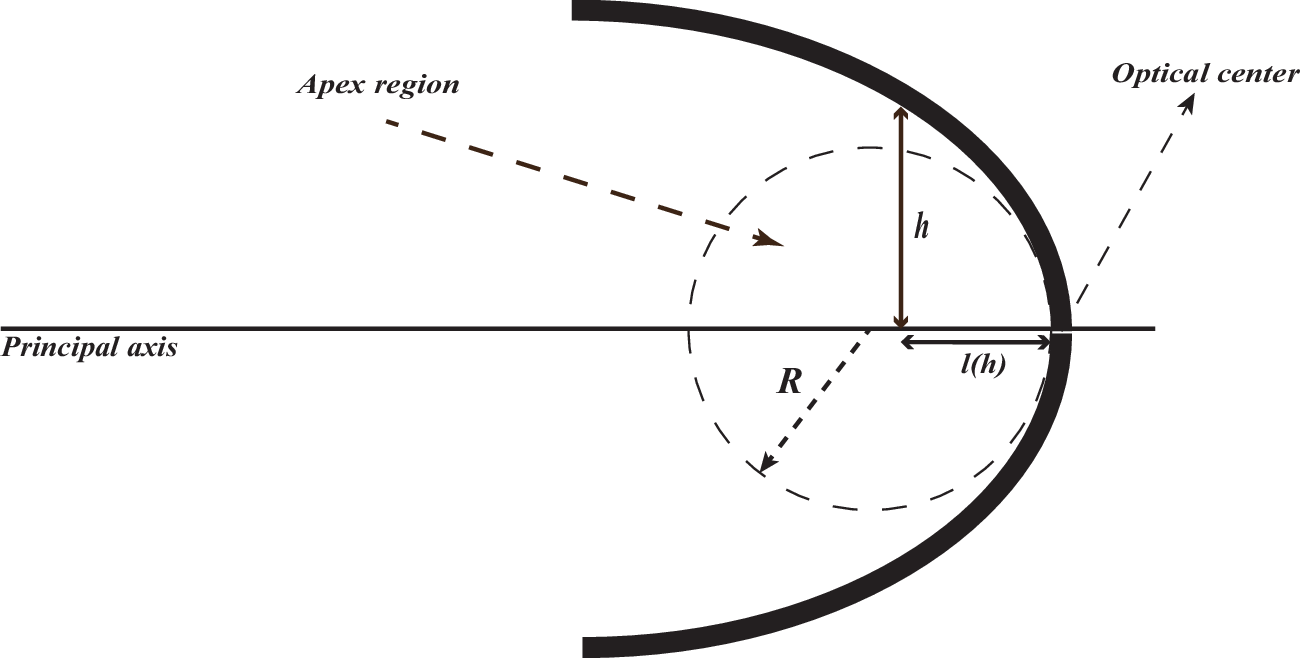}
\caption{Schematic diagram of a conic section. The area inside the dashed circle is the apex region of a conic section.} \label{curved-mirror-3}
\end{center}
\end{figure}

The general mathematical description of the two dimensional conic sections~\cite{Watson99,Baker43,Gantinel01} is given by
\begin{equation}
(1 - e^2)l^2 + h^2 - 2Rl = 0,
\label{general-conic}
\end{equation}
where $l$ is the horizontal distance from the optical center of the mirror system, $h$ is the vertical distance from the principal axis of the mirror system, $e$ is the eccentricity of the conic section, and $R$ is the apical radius of curvature (apex), given by
\begin{equation}
 R = \left\{\frac{\left[1+\left(\frac{dl}{dh}\right)^2\right]^\frac{3}{2}}{\left|\frac{d^2l}{dh^2}\right|}\right\}_{h=0}
\label{apex-curvature}
\end{equation}
The apex region of a given conic section is illustrated by Figure~\ref{curved-mirror-3}.

Eq.~(\ref{general-conic}) is known as {\it Baker's equation}~\cite{Baker43,Gantinel01}. Solving Eq.~(\ref{general-conic}) for the horizontal distance $l$, we get
\begin{equation}
l_{\pm}(h)=\pm \left(\frac{R-\sqrt{R^2-(1-e^2)h^2}}{1-e^2}\right).
\label{general-conic-2}
\end{equation}
It is clear from Eq.~(\ref{general-conic-2}) that $l(h)$ is an even function of $h$, with $l(0) = 0$ due to the fact that the $l(h)$ vanishes at the optical center. The $\pm$ signs refer to concave / convex mirror systems.
There are four types of mirrors based on the variation of eccentricity $e$ :
\begin{itemize}
\item Spherical mirror ($e = 0$).
\item Elliptic mirror ($0 < e < 1$).
\item Parabolic mirror ($e = 1$).
\item Hyperbolic mirror ($e > 1$).
\end{itemize}

\section{Applications of Fermat's principle on a static curved mirror}
\label{static-mirror}

\begin{figure}
\begin{center}
\includegraphics[scale=0.5]{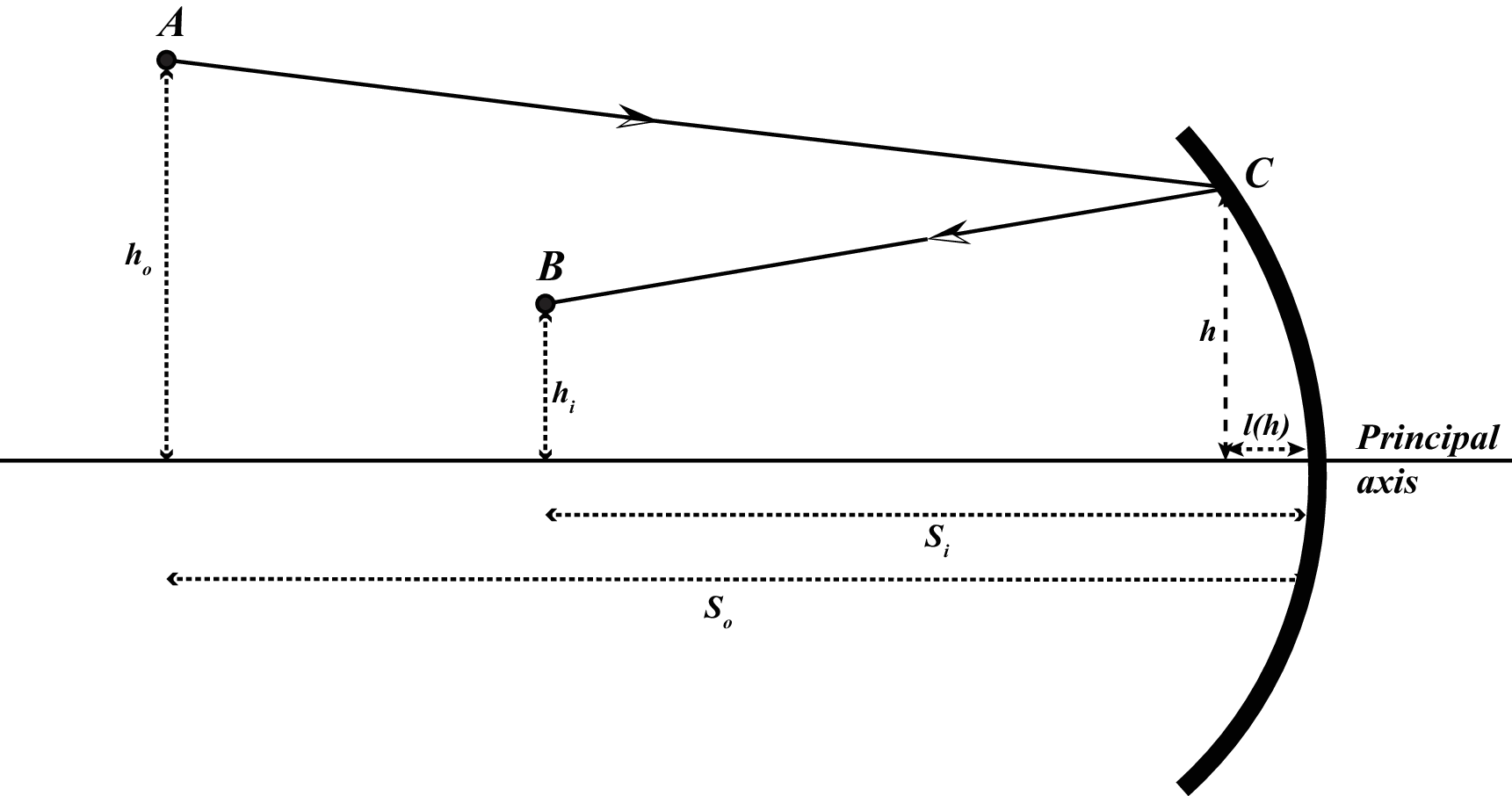}
\caption{Schematic diagram of a static curved mirror system. The ($S_0$,$h_0$) and ($S_i$,$h_i$) are sets of coordinates related to the object and its associated image, respectively. The solid line {\bf A-C-B} is the optical path length (OPL) of the light.} \label{curved-mirror}
\end{center}
\end{figure}

Figure~\ref{curved-mirror} shows the propagation of light from point {\bf A} to point {\bf B}, with reflection of light at point {\bf C} on the surface of the mirror.

The {\it optical path length} (OPL) of the light from {\bf A} to {\bf B} is
\begin{equation}
\mbox{OPL} (h) = S_0 \sqrt{\left[1 - \frac{l(h)}{S_0}\right]^2 + \left[\frac{h - h_0}{S_0}\right]^2} + S_i \sqrt{\left[1 - \frac{l(h)}{S_i}\right]^2 + \left[\frac{h_i - h}{S_i}\right]^2}.
\label{OPL-static}
\end{equation}
The travelling time of light can be obtained by dividing the OPL~(\ref{OPL-static}) with the speed of light $c$ of the corresponding medium :
\begin{equation}
t(h) = \frac{S_0}{c} \sqrt{\left[1 - \frac{l(h)}{S_0}\right]^2 + \left[\frac{h - h_0}{S_0}\right]^2} + \frac{S_i}{c} \sqrt{\left[1 - \frac{l(h)}{S_i}\right]^2 + \left[\frac{h_i - h}{S_i}\right]^2}.
\label{time-static}
\end{equation}
The application of Fermat's principle can be accomplished by setting the first derivative of Eq.~(\ref{time-static}) to zero. The result is
\begin{eqnarray}
\frac{dt}{dh} (h) & = & \frac{\frac{h - h_0}{S_0}-\left(1-\frac{l(h)}{S_0}\right)\frac{dl}{dh}}{c \sqrt{\left[1 - \frac{l(h)}{S_0}\right]^2 + \left[\frac{h - h_0}{S_0}\right]^2}} - \frac{\frac{h_i - h}{S_i}+\left(1-\frac{l(h)}{S_i}\right)\frac{dl}{dh}}{c \sqrt{\left[1 - \frac{l(h)}{S_i}\right]^2 + \left[\frac{h_i - h}{S_i}\right]^2}} \nonumber \\
& = & 0.
\label{Fermat-static}
\end{eqnarray}
Since it is very difficult to solve Eq.~(\ref{Fermat-static}) for $h$, we shall analyze it by taking the Taylor expansion of Eq.~(\ref{Fermat-static}) near $h = 0$ :
\begin{equation}
\frac{dt}{dh}(h) = \sum_{n=1}^{\infty} \frac{1}{(n-1)!}\frac{d^n t}{dh^n}(0)h^{n-1} = 0,
\label{Taylor-expansion}
\end{equation}
leading to
\begin{equation}
\frac{d^n t}{dh^n}(0)= 0, \hspace{5mm} n=1, 2, 3, ...
\label{Taylor-expansion-2}
\end{equation}

\subsection{The Gaussian formula for a static mirror}
\label{gaussian-static}

The first two terms of Eq.~(\ref{Taylor-expansion-2}) after the insertion of Eq.~(\ref{time-static}) lead to
\begin{equation}
\frac{\left(1-\frac{l(0)}{S_0}\right)\dot{l}(0)+\frac{h_0}{S_0}}{\sqrt{\left(1-\frac{l(0)}{S_0}\right)^2+\frac{h_0^2}{S_0^2}}}
+\frac{\left(1-\frac{l(0)}{S_i}\right)\dot{l}(0)+\frac{h_i}{S_i}}{\sqrt{\left(1-\frac{l(0)}{S_i}\right)^2+\frac{h_i^2}{S_i^2}}}=0,
\label{Taylor-1-static}
\end{equation}
and
\begin{eqnarray}
& & \frac{\frac{1}{S_0}\left(1+\dot{l}^2(0)\right)-\left(1-\frac{l(0)}{S_0}\right)\ddot{l}(0)}{\sqrt{\left(1-\frac{l(0)}{S_0}\right)^2+\frac{h_0^2}{S_0^2}}}- \frac{\frac{1}{S_0}\left(\left(1-\frac{l(0)}{S_0}\right)\dot{l}(0)+\frac{h_0}{S_0}\right)^2}{\left(\left(1-\frac{l(0)}{S_0}\right)^2+\frac{h_0^2}{S_0^2}\right)^\frac{3}{2}} \nonumber \\
& + & \frac{\frac{1}{S_i}\left(1+\dot{l}^2(0)\right)-\left(1-\frac{l(0)}{S_i}\right)\ddot{l}(0)}{\sqrt{\left(1-\frac{l(0)}{S_i}\right)^2+\frac{h_i^2}{S_i^2}}}- \frac{\frac{1}{S_i}\left(\left(1-\frac{l(0)}{S_i}\right)\dot{l}(0)+\frac{h_i}{S_i}\right)^2}{\left(\left(1-\frac{l(0)}{S_i}\right)^2+\frac{h_i^2}{S_i^2}\right)^\frac{3}{2}} \nonumber \\
& & = 0,
\label{Taylor-2-static}
\end{eqnarray}
where $\dot{l}(h)$ and $\ddot{l}(h)$ are the first and the second derivatives of $l(h)$ with respect to $h$, respectively.

In order to get a clear interpretation of these terms, we consider two usual assumptions in geometrical optics :
\begin{itemize}
\item It is common to use the so-called {\it paraxial ray approximation}, i.e. when the rays are at small angles with respect to the principal axis. The consequences of the approximation are
    \begin{equation}
    \left|\frac{h_0}{S_0}\right| \ll 1, \hspace{5mm} \left|\frac{h_i}{S_i}\right| \ll 1,
    \label{paraxial-condition}
    \end{equation}
    i.e. we may neglect the higher order contributions of $\frac{h_0}{S_0}$ and $\frac{h_i}{S_i}$ on Eq.~(\ref{Taylor-expansion-2}).
\item It is customary to consider mirrors whose curvatures are symmetric with respect to their associated principal axes. Consequently, the function $l(h)$ must be an even function: $l(-h)=l(h)$, and all odd derivatives of $l(h)$ vanish at $h=0$~\cite{Lemons94}. The shapes of mirrors discussed in Section~\ref{conic-section} satisfy this condition.
\end{itemize}
Based on these assumptions, Eqs.~(\ref{Taylor-1-static}) and (\ref{Taylor-2-static}) lead to
\begin{equation}
\frac{h_0}{S_0}+\frac{h_i}{S_i}=0 \hspace{5mm} \Longrightarrow \hspace{5mm} \frac{h_i}{h_0}=-\frac{S_i}{S_0},
\label{Taylor-3-static}
\end{equation}
\begin{equation}
\frac{1}{S_0}+\frac{1}{S_i}=2\ddot{l}(0).
\label{Taylor-4-static}
\end{equation}
Eq.~(\ref{Taylor-3-static}) expresses the usual lateral magnification formula, and Eq.~(\ref{Taylor-4-static}) is the Gaussian formula relating the object distance $S_0$ to the image distance $S_i$, with the focal length $f$ defined as
\begin{equation}
f \equiv \frac{1}{2 \ddot{l}(0)}.
\label{focal-static}
\end{equation}
Using the horizontal distance given by Eq.~(\ref{general-conic-2}), it is interesting to note that the second derivative $\ddot{l}(0) = \pm \frac{1}{R}$ regardless of the eccentricity $e$. Hence the focal length for any shape of mirror generated by a conic section is
\begin{equation}
f_{\pm} = \pm \frac{R}{2}.
\label{focal-conic-static}
\end{equation}
If the conic section is a circle, Eq.~(\ref{focal-conic-static}) gives the usual focal length of a spherical mirror, since the apex of a circle is equal to its radius. In general, Eq.~(\ref{focal-conic-static}) is equally true for any shape of mirror generated by a conic section, where $R$ is generalized to be the apex of the associated conic section.

\subsection{Spherical aberrations in static mirrors}
\label{aberration-static}

\begin{figure}
\begin{center}
\includegraphics[scale=0.5]{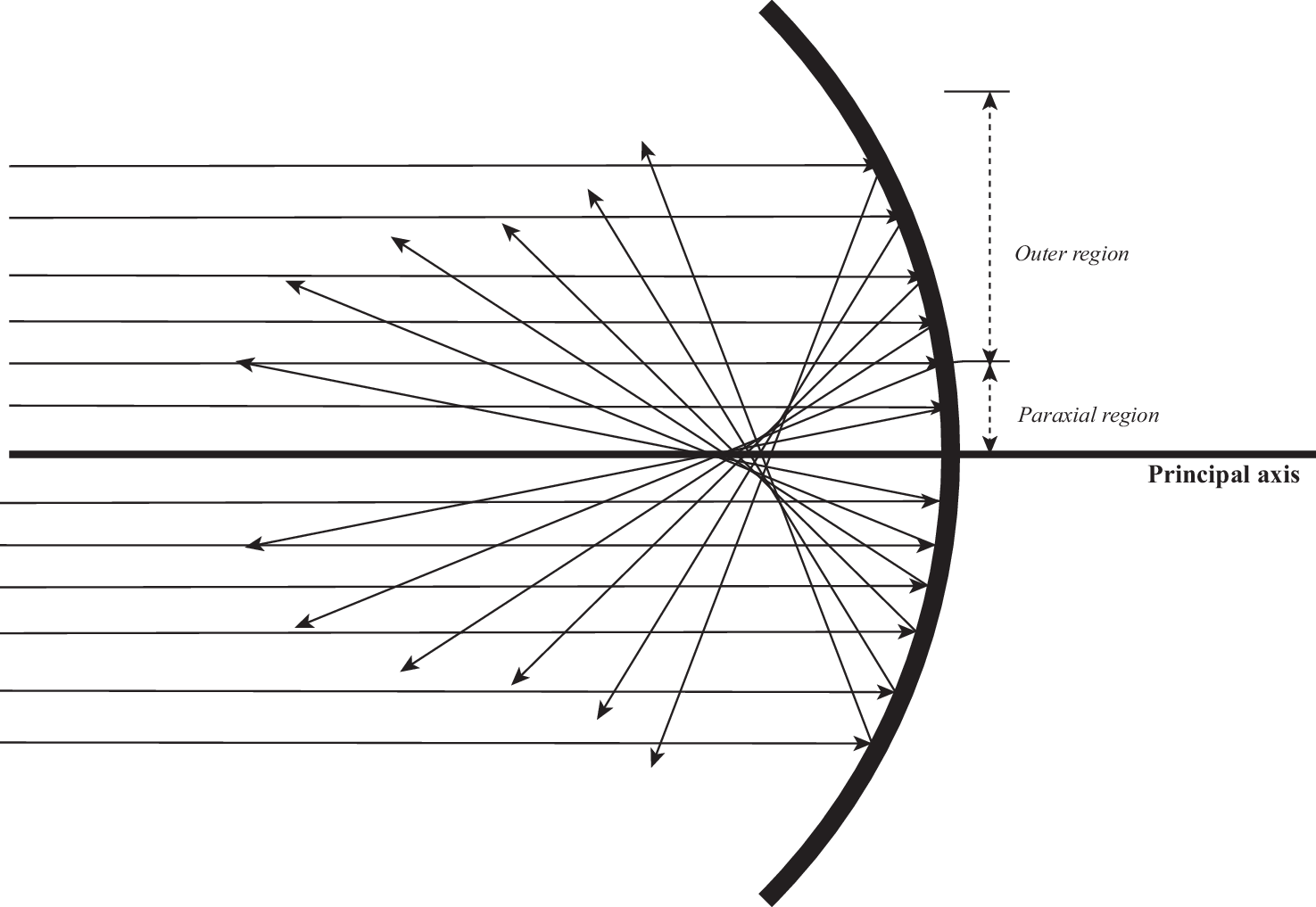}
\caption{Schematic diagram of spherical abberation.} \label{aberration}
\end{center}
\end{figure}

Spherical aberration occurs in the curved mirror system when for incident light rays parallel to the principal axis, the reflected light rays coming from the outer region of the mirror do not arrive at the same point as those from the paraxial region~\cite{Hecth02}. The schematic diagram of the phenomenon is shown in Figure~\ref{aberration}.

 In the paraxial region, the Gaussian formula~(\ref{Taylor-4-static}) holds. We may obtain the focal length of the curved mirror by setting $S_0 \rightarrow \infty$, i.e. when the light source is relatively far away from the mirror, and the distance of the associated image $S_i$ is equal to the focal length. By Eq.~(\ref{Taylor-3-static}), the image height $h_i = 0$ regardless of the object height $h_0$. As we consider light rays coming from the outer region, we have to start from Eq.~(\ref{Fermat-static}), take $S_0 \rightarrow \infty$ and $h_i =0$, and solve the resulting equation for $S_i$. The result is
\begin{equation}
S_i(h)= f(h) = l(h) +\frac{h\left(1-\dot{l}^2(h)\right)}{2 \dot{l}(h)}.
\label{aberration-static-1}
\end{equation}
Inserting Eq.~(\ref{general-conic-2}) into Eq.~(\ref{aberration-static-1}), we obtain
\begin{equation}
f(h) = \pm \left(\frac{R-\sqrt{R^2-(1-e^2)h^2}}{1-e^2} + \frac{R^2 - (2 - e^2)h^2}{2 \sqrt{R^2-(1-e^2)h^2}}\right)
\label{aberration-static-conic}
\end{equation}
It is clear from Eq.~(\ref{aberration-static-conic}) that in general, the intersection of the reflected rays to the principal axis depends on the vertical distance $h$ and the shape of mirror.

\begin{figure}
\begin{center}
\includegraphics[scale=0.5]{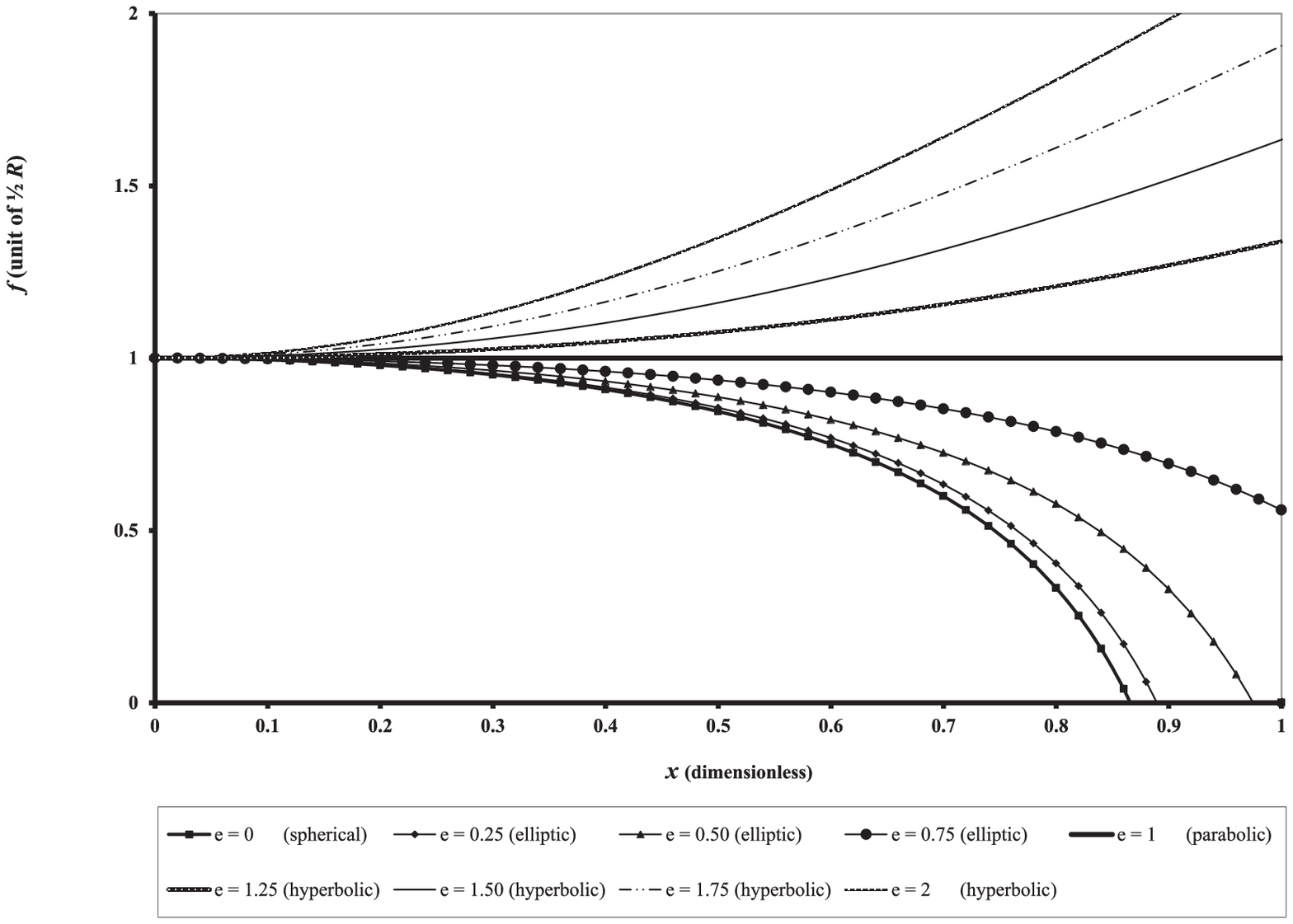}
\caption{The spherical aberration of concave mirrors of some conic sections based on Eq.~(\ref{aberration-static-conic-x}).} \label{aberration-static-graph}
\end{center}
\end{figure}

To obtain a clearer analysis of the spherical aberration, we introduce a dimensionless quantity $x$ given by
\begin{equation}
x \equiv \frac{h}{R}.
\label{x-substitute}
\end{equation}
Using the $h \longrightarrow x$ substitution, Eq.~(\ref{aberration-static-conic}) becomes
\begin{eqnarray}
f(h) \longrightarrow f(x) = & \pm \frac{R}{2} & \left[\frac{2 \left(1-\sqrt{1-(1-e^2) x^2}\right)}{1 - e^2} \right. \nonumber \\
&  & \left. + \frac{1 - (2-e^2) x^2}{\sqrt{1-(1-e^2) x^2}} \right].
\label{aberration-static-conic-x}
\end{eqnarray}
Figure~\ref{aberration-static-graph} shows the plot of $f(x)$ with respect to $x$ for selected conic sections.

Now we may analyze the behaviour of the spherical aberration based on Eq.~(\ref{aberration-static-conic-x}) :
\begin{enumerate}
\item The Taylor expansion of Eq.~(\ref{aberration-static-conic-x}) in powers of $x$ gives
\begin{eqnarray}
f(x) = & \pm \frac{R}{2} & \left(1 - \frac{(1 - e^2)x^2}{2} - \frac{(1 - e^2)(3 - e^2)x^4}{8} \right. \nonumber \\
&  & \left. - \frac{(5 - e^2)(1 - e^2)^2 x^6}{16} + O(x^8)\right).
\label{aberration-static-conic-expansion}
\end{eqnarray}
The expansion~(\ref{aberration-static-conic-expansion}) gives the same results as obtained by Watson using a different approach~\cite{Watson99}. Eq.~(\ref{aberration-static-conic-expansion}) indicates that the intersection point of the reflected ray (or the extension of the reflected ray in the case of convex mirrors) with the principal axis is coming closer to the optical center for mirror shapes with eccentricity $0 \leq e < 1$ as $x$ increases, and is moving away from the optical center for hyperbolic mirrors ($e > 1$). For the parabolic mirror ($e = 1$), all the reflected rays (or the extension of the reflected ray in the case of convex mirrors) come to a single point whose horizontal distance is $\pm \frac{R}{2}$ from the optical center. This means that the parabolic mirror does not suffer spherical aberration, as expected.

\item If we are considering the paraxial region where $ h \ll R$ (or where $x$ is sufficiently small), we may neglect the higher order terms in Eq.~(\ref{aberration-static-conic-expansion}), leaving $f(x) \approx \pm \frac{R}{2}$. Consequently, the absolute error $\delta_f (x)$ generated from the approximation is given by
    \begin{eqnarray}
    \delta_f (x) & \equiv & \left|f(x)-\frac{R}{2}\right| \nonumber \\
    & = & \frac{R}{2} \left| \frac{1-(2-e^2) x^2}{\sqrt{1 - (1-e^2)x^2}}+\frac{2 \left(1-\sqrt{1-(1-e^2)x^2} \right)}{1-e^2}-1 \right|.
    \label{relative-error-static}
    \end{eqnarray}
    The growth of the absolute error with respect to $x$ for some concave mirror is given by Figure~\ref{error-static}.
    \begin{figure}
    \begin{center}
    \includegraphics[scale=0.5]{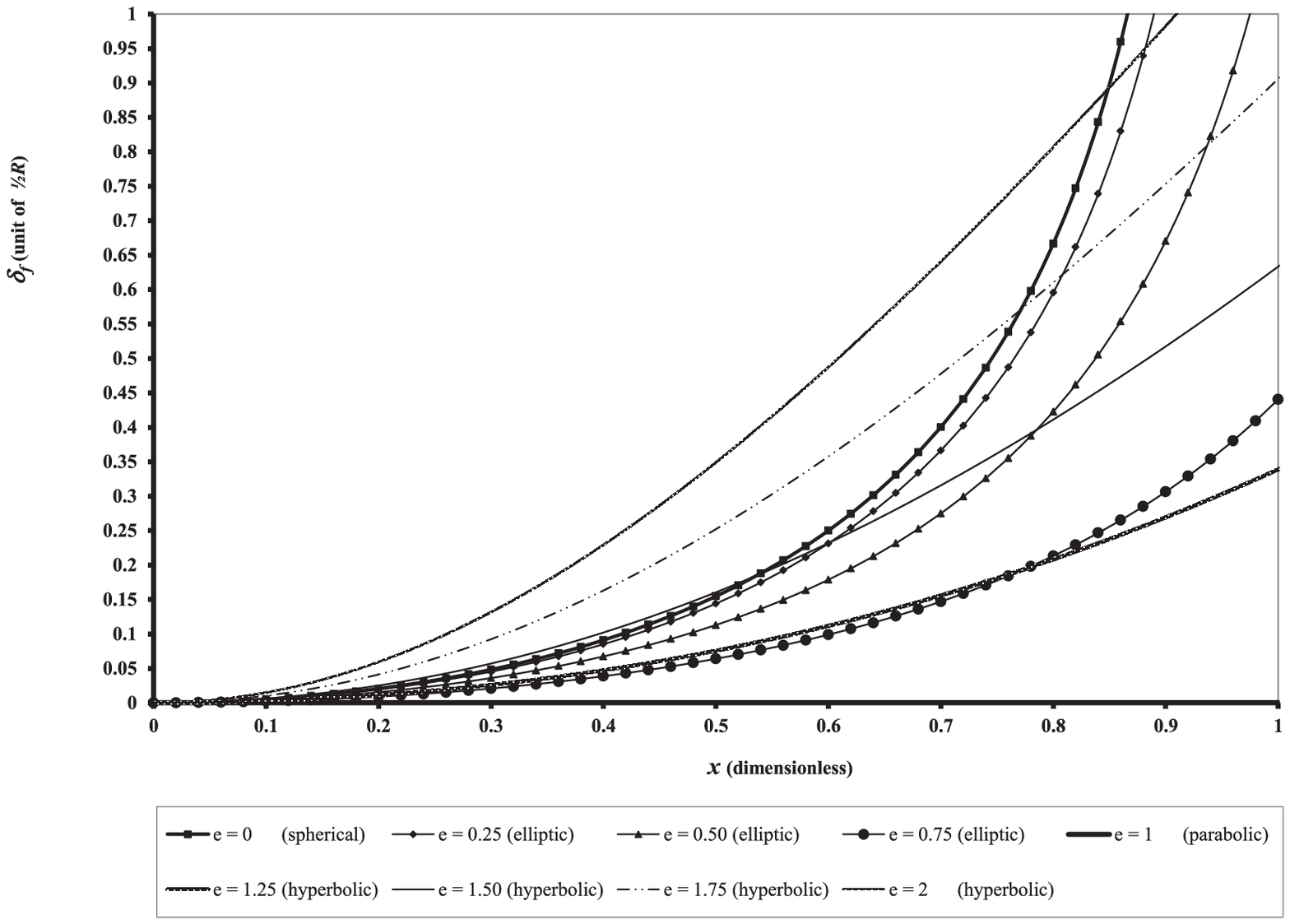}
    \caption{The plot of relative error $\delta_f(x)$ with respect to $x$ of concave mirrors of some conic sections based on Eq.~(\ref{relative-error-static}).} \label{error-static}
    \end{center}
    \end{figure}
    To obtain the paraxial range of $f(x)$, it is customary to use the acceptable error of $5\% \times \frac{R}{2} = 0.025 R$. Using Figure~\ref{error-static}, we may observe errors of $\delta_f \leq 0.025 R$ at $x < x_0$, where
    \begin{enumerate}
    \item $x_0 \approx 0.3$ for spherical mirrors,
    \item $x_0 > 0.3$ for elliptic mirrors,
    \item $x_0 > 0.2$ for hyperbolic mirrors with eccentricity within $1 < e \leq 2$.
    \end{enumerate}
    In fact, solving Eq.~(\ref{relative-error-static}) for $\delta_f (x) \leq 0.025 R$ numerically leads to the more precise results of
    \begin{enumerate}
    \item $x_0 \approx 0.304911$ for spherical mirrors,
    \item $x_0 > 0.304911$ for elliptic mirrors,
    \item $x_0 > 0.183379$ for hyperbolic mirrors with eccentricity within $1 < e \leq 2$.
    \end{enumerate}
    It is also interesting to see from Figure~\ref{error-static} that the errors grow rapidly at some point beyond $x \approx 0.6$ for spherical and elliptic mirrors. The errors for hyperbolic mirrors also grow, but not as rapidly as the spherical and elliptic cases. \label{paraxial}

\item From (\ref{paraxial}) above, it is interesting to note that the spherical aberration may be reduced by constructing the mirrors generated by conic sections with eccentricity close to 1. It is also practical to reduce the spherical aberration by setting a sufficiently small value of $x$, i.e. by taking a sufficiently large value of the apical radius $R$ compared to $h$.
\end{enumerate}

\section{Applications of Fermat's principle on a relativistic curved mirror}
\label{moving-mirror}

\begin{figure}
\begin{center}
\includegraphics[scale=0.5]{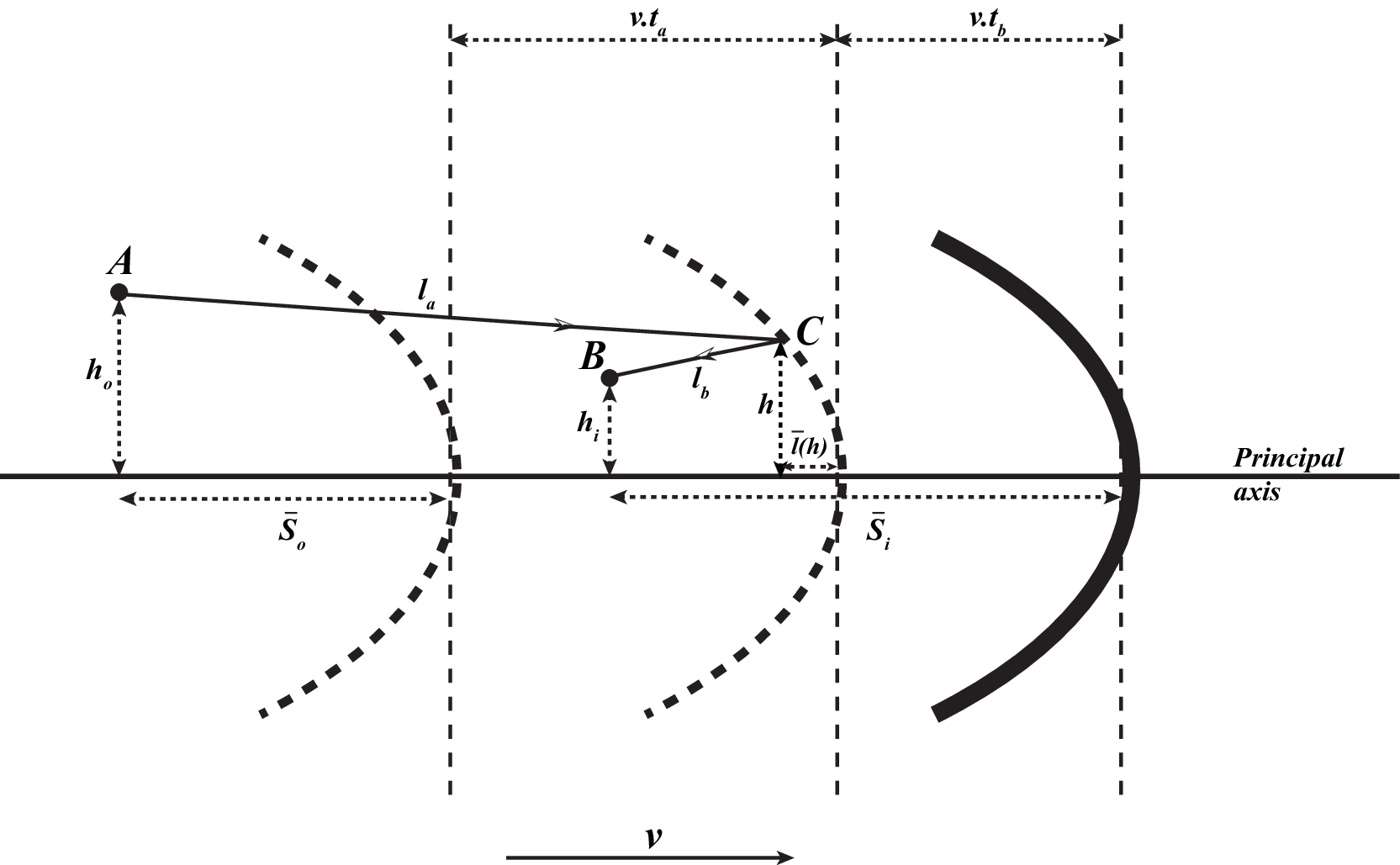}
\caption{Schematic diagram of the relativistic curved mirror system. The OPL from {\bf A} to {\bf B} is shown by solid lines.} \label{curved-mirror-moving}
\end{center}
\end{figure}

Now let us study the behaviour of a mirror system where the object, the image and the curved mirror are all moving parallel to the principal axis at a constant speed $v$ to the right relative to the observer, as shown in Figure~\ref{curved-mirror-moving}. The light ray travels from point {\bf A} at time $t = 0$, is reflected at point {\bf C} on the surface of the mirror at time $t = t_a$, and finally reaches point {\bf B} at time $t = t_a + t_b$. During the movement of the mirror system, it is important to keep in mind two important rules :
\begin{itemize}
\item the speed of light {\it remains the same} as observed by either the rest or moving observers, as stated by the second postulate of the special theory of relativity~\cite{Einstein05}.
\item lengths that are parallel to the direction of the movement appear shorter with respect to the rest observer due to the Lorentz contraction.
\end{itemize}

Based on the two rules above, we construct the OPL of the light from {\bf A} to {\bf C} and from {\bf C} to {\bf B} as
\begin{equation}
l_a (h) = \bar{S}_0 \sqrt{\left[1 + \frac{v t_a - \bar{l}(h)}{\bar{S}_0}\right]^2 + \left[\frac{h - h_0}{\bar{S}_0}\right]^2},
\label{OPL-moving-a}
\end{equation}
and
\begin{equation}
l_b (h) = \bar{S}_i \sqrt{\left[1 - \frac{v t_b + \bar{l}(h)}{\bar{S}_i}\right]^2 + \left[\frac{h_i - h}{\bar{S}_i}\right]^2},
\label{OPL-moving-b}
\end{equation}
respectively. The $\bar{S}_0, \bar{S}_i$ and $\bar{l}(h)$ are the Lorentz contracted version of $S_0$, $S_i$ and $l(h)$, respectively, i.e.
\begin{eqnarray}
\bar{S}_0 & = & \frac{S_0}{\gamma}, \nonumber \\
\bar{S}_i & = & \frac{S_i}{\gamma}, \nonumber\\
\bar{l}(h) & = & \frac{l(h)}{\gamma},
\label{Lorentz}
\end{eqnarray}
where $\gamma \equiv \left(1 - \frac{v^2}{c^2}\right)^{-\frac{1}{2}}$.

The travelling time from {\bf A} to {\bf C} can be obtained from Eq.~(\ref{OPL-moving-a}) by dividing it with the speed of light $c$ and solving the resulting equation for $t_a$. The result is
\begin{equation}
t_a (h) = \frac{\bar{S}_0}{c} \gamma^2 \left[\sqrt{\left(1 - \frac{\bar{l}(h)}{\bar{S}_0}\right)^2 + \frac{1}{\gamma^2}\left(\frac{h - h_0}{\bar{S}_0}\right)^2} + \beta \left(1-\frac{\bar{l}(h)}{\bar{S}_0}\right) \right],
\label{time-moving-a}
\end{equation}
where $\beta \equiv \frac{v}{c}$.
Similarly, the travelling time from {\bf C} to {\bf B} can be obtained from Eq.~(\ref{OPL-moving-b}), leading to
\begin{equation}
t_b (h) = \frac{\bar{S}_i}{c} \gamma^2 \left[\sqrt{\left(1 - \frac{\bar{l}(h)}{\bar{S}_i}\right)^2 + \frac{1}{\gamma^2}\left(\frac{h_i - h}{\bar{S}_i}\right)^2} - \beta \left(1-\frac{\bar{l}(h)}{\bar{S}_i}\right) \right].
\label{time-moving-b}
\end{equation}
The total travelling time is then $t(h)=t_a (h) + t_b (h)$.

Applying Fermat's principle by setting the first derivative of the total travelling time $t(h)$ to zero, we obtain
\begin{equation}
\frac{\frac{h - h_0}{\gamma^2 \bar{S}_0}-\left(1 - \frac{\bar{l}(h)}{\bar{S}_0}\right)\dot{\bar{l}}(h)}{\sqrt{\left(1 - \frac{\bar{l}(h)}{\bar{S}_0}\right)^2 + \frac{1}{\gamma^2}\left(\frac{h - h_0}{\bar{S}_0}\right)^2}} - \frac{\frac{h_i - h}{\gamma^2 \bar{S}_i}+\left(1 - \frac{\bar{l}(h)}{\bar{S}_i}\right)\dot{\bar{l}}(h)}{\sqrt{\left(1 - \frac{\bar{l}(h)}{\bar{S}_i}\right)^2 + \frac{1}{\gamma^2}\left(\frac{h_i - h}{\bar{S}_i}\right)^2}} = 0.
\label{Fermat-moving}
\end{equation}
For $v \ll c$, Eq.~(\ref{Fermat-moving}) leads to Eq.~(\ref{Fermat-static}). As in the static case, Eq.~(\ref{Fermat-moving}) is very difficult to solve for $h$, and we shall use the Taylor expansion~(\ref{Taylor-expansion}) and the relation~(\ref{Taylor-expansion-2}) again to analyze it.

\subsection{The Gaussian formula for the relativistic mirrors}
\label{gaussian-moving}

The first two terms of the relation~(\ref{Taylor-expansion-2}) obtained from Eq.~(\ref{Fermat-moving}) give
\begin{equation}
\frac{\frac{h_0}{\gamma^2 \bar{S}_0}-\left(1 - \frac{\bar{l}(0)}{\bar{S}_0}\right)\dot{\bar{l}}(0)}{\sqrt{\left(1 - \frac{\bar{l}(0)}{\bar{S}_0}\right)^2 + \frac{1}{\gamma^2}\left(\frac{h_0}{\bar{S}_0}\right)^2}} + \frac{\frac{h_i}{\gamma^2 \bar{S}_i}+\left(1 - \frac{\bar{l}(0)}{\bar{S}_i}\right)\dot{\bar{l}}(0)}{\sqrt{\left(1 - \frac{\bar{l}(0)}{\bar{S}_i}\right)^2 + \frac{1}{\gamma^2}\left(\frac{h_i}{\bar{S}_i}\right)^2}} = 0,
\label{Taylor-1-moving}
\end{equation}
\begin{eqnarray}
& & \frac{\frac{1}{\bar{S}_0}\left(\frac{1}{\gamma^2}+\dot{\bar{l}}^2(0)\right)-\left(1-\frac{\bar{l}(0)}{\bar{S}_0}\right)\ddot{\bar{l}}(0)}{\sqrt{\left(1-\frac{\bar{l}(0)}{\bar{S}_0}\right)^2+\frac{h_0^2}{\gamma^2 \bar{S}_0^2}}}- \frac{\frac{1}{\bar{S}_0}\left(\left(1-\frac{\bar{l}(0)}{\bar{S}_0}\right)\dot{\bar{l}}(0)+\frac{h_0}{\gamma^2 \bar{S}_0}\right)^2}{\left(\left(1-\frac{\bar{l}(0)}{\bar{S}_0}\right)^2+\frac{h_0^2}{\gamma^2 \bar{S}_0^2}\right)^\frac{3}{2}} \nonumber \\
& + & \frac{\frac{1}{\bar{S}_i}\left(\frac{1}{\gamma^2}+\dot{\bar{l}}^2(0)\right)-\left(1-\frac{\bar{l}(0)}{\bar{S}_i}\right)\ddot{\bar{l}}(0)}{\sqrt{\left(1-\frac{\bar{l}(0)}{\bar{S}_i}\right)^2+\frac{h_i^2}{\gamma^2 \bar{S}_i^2}}}- \frac{\frac{1}{\bar{S}_i}\left(\left(1-\frac{\bar{l}(0)}{\bar{S}_i}\right)\dot{\bar{l}}(0)+\frac{h_i}{\gamma^2 \bar{S}_i}\right)^2}{\left(\left(1-\frac{\bar{l}(0)}{\bar{S}_i}\right)^2+\frac{h_i^2}{\gamma^2 \bar{S}_i^2}\right)^\frac{3}{2}} = 0.
\label{Taylor-2-moving}
\end{eqnarray}
Since $\bar{l}(h)$, like $l(h)$, is an even function of $h$, and using the paraxial ray approximation~(\ref{paraxial-condition}), Eqs~(\ref{Taylor-1-moving}) and (\ref{Taylor-2-moving}) lead to
\begin{equation}
\frac{1}{\gamma^2} \left(\frac{h_0}{\bar{S}_0}+\frac{h_i}{\bar{S}_i}\right)=0 \hspace{5mm} \Longleftrightarrow \hspace{5mm} \frac{h_i}{h_0} = -\frac{\bar{S}_i}{\bar{S}_0} = -\frac{S_i}{S_0},
\label{Taylor-3-moving}
\end{equation}
\begin{equation}
\frac{1}{\bar{S}_0}+\frac{1}{\bar{S}_i}=2 \gamma^2 \ddot{\bar{l}}(0).
\label{Taylor-4-moving}
\end{equation}

Comparing Eqs.~(\ref{Taylor-3-static}) and (\ref{Taylor-3-moving}), it is clear that the lateral magnification formula is preserved, i.e. the magnification rule appears the same in either static or relativistic case. This is due to the fact that any length in the direction perpendicular to the direction of the motion does not suffer Lorentz contraction. The Gaussian formula, however, suffers a modification of the focal length of the mirror, i.e.
\begin{equation}
f \equiv \frac{1}{2 \ddot{l}(0)} \Longrightarrow \bar{f} \equiv \frac{1}{2 \gamma^2 \ddot{\bar{l}}(0)} = \frac{f}{\gamma},
\label{focal-moving}
\end{equation}
based on comparison between Eqs.~(\ref{Taylor-4-static}) and (\ref{Taylor-4-moving}). Hence, the focal length of the relativistic mirror is Lorentz contracted for the rest observer. Inserting Eq.~(\ref{focal-conic-static}) into Eq.~(\ref{focal-moving}), we obtain the focal length of the relativistic curved mirror with the shape of a conic section as
\begin{equation}
\bar{f} = \frac{R}{2 \gamma}.
\label{focal-moving-conic}
\end{equation}
Due to the quadratic term of the speed $v$ in $\gamma$, the Gaussian formula~(\ref{Taylor-4-moving}) and the focal length~(\ref{focal-moving}) are independent of the direction of movement.

\subsection{Spherical aberrations in relativistic mirrors}
\label{aberration-moving}

Using the same arguments as used in the discussion of spherical aberration in Section~\ref{aberration-static}, we shall start from Eq.~(\ref{Fermat-moving}), take $\bar{S}_0 \rightarrow \infty$ and $h_i =0$, and solve the resulting equation for $\bar{S}_i$. The result is
\begin{eqnarray}
\bar{S}_i(h)= \bar{f}(h) & = & \bar{l}(h) +\frac{h\left(\frac{1}{\gamma^2}-\dot{\bar{l}}^2(h)\right)}{2 \dot{\bar{l}}(h)} \nonumber \\
& = & \frac{1}{\gamma} \left(l(h) +\frac{h\left(1-\dot{l}^2(h)\right)}{2 \dot{l}(h)}\right) = \frac{f(h)}{\gamma}. \label{aberration-moving-1}
\end{eqnarray}
Eq.~(\ref{aberration-moving-1}) indicates that the spherical aberration relation (\ref{aberration-static-conic}) is Lorentz contracted, which is consistent with Eq.~(\ref{focal-moving}). Again, Eq.~(\ref{aberration-moving-1}) is independent of the direction of movement due to the quadratic term of the speed $v$ in $\gamma$.

In order to analyze further the effect of speed to the spherical aberration, we use the $h \longrightarrow x$ substitution introduced by Eq.~(\ref{x-substitute}) in Section~\ref{aberration-static}. Inserting Eq.~(\ref{general-conic-2}) into Eq.~(\ref{aberration-moving-1}) and applying the $h \longrightarrow x$ substitution, we obtain
\begin{eqnarray}
\bar{f}(x) & = &  \pm \frac{R}{2 \gamma} \left[\frac{2 \left(1-\sqrt{1-(1-e^2) x^2}\right)}{1 - e^2} + \frac{1 - (2-e^2) x^2}{\sqrt{1-(1-e^2) x^2}} \right].
\label{aberration-moving-x}
\end{eqnarray}
Let us analyze the effect of the speed to the spherical aberration based on Eq.~(\ref{aberration-moving-x}) :
\begin{enumerate}
\item Figure~\ref{aberration-moving-spherical} shows the spherical aberration of a relativistic spherical concave mirror for the purpose of illustration of Eq.~(\ref{aberration-moving-x}). It is clear from the figure that the reflection to the optical center of the spherical mirror ($e = 0$) occurs at $h > 0.8 R$ regardless of the speed of the mirror system. By solving $\bar{f}(x) = 0$ analytically, we obtain the more precise result of $h = \frac{\sqrt{3}}{2} R \approx 0.866025 R$. The fact that this value of $h$ is independent of the speed can be understood from the form of Eq~(\ref{aberration-moving-1}), where the factor $\bar{f}(x)$ is a multiple of $f(x)$ by the factor $\frac{1}{\gamma}$.
    \begin{figure}
    \begin{center}
    \includegraphics[scale=0.5]{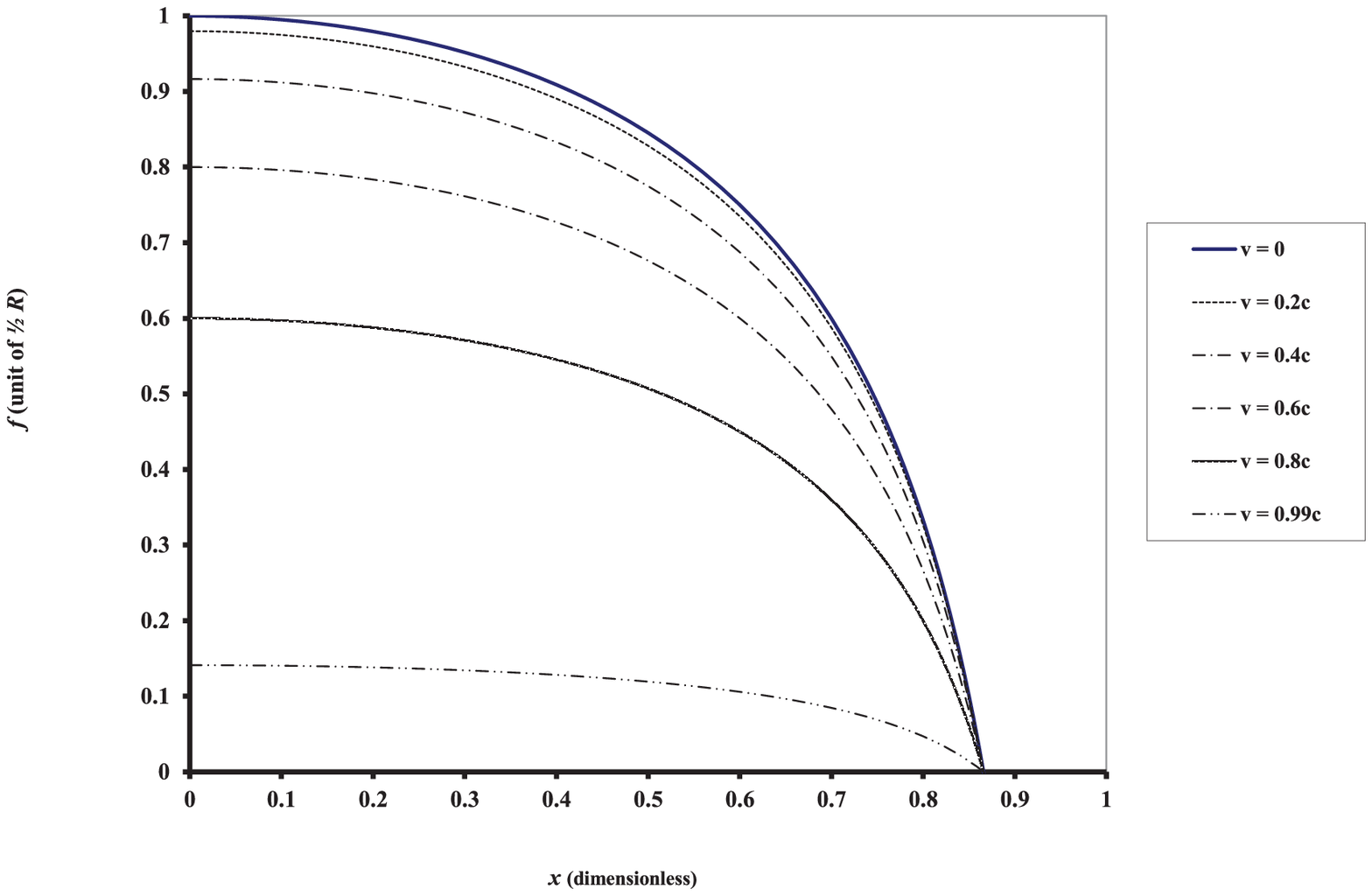}
    \caption{The spherical aberration of the relativistic spherical concave mirror ($e = 0$) based on Eq.~(\ref{aberration-moving-x}).} \label{aberration-moving-spherical}
    \end{center}
    \end{figure}

\item The Taylor expansion of Eq.(\ref{aberration-moving-x}) in powers of $x$ gives
    \begin{eqnarray}
    \bar{f}(x) = & \pm \frac{R}{2 \gamma} & \left(1 - \frac{(1 - e^2)x^2}{2} - \frac{(1 - e^2)(3 - e^2)x^4}{8} \right. \nonumber \\
    &  & \left. - \frac{(5 - e^2)(1 - e^2)^2 x^6}{16} + O(x^8)\right).
    \label{aberration-moving-conic-expansion}
    \end{eqnarray}
    Within the paraxial region where $h \ll R$, we may again neglect the higher order terms in Eq.~(\ref{aberration-moving-conic-expansion}) so $\bar{f}(x) \approx \pm \frac{R}{2 \gamma}$. Using Eq.~(\ref{aberration-moving-x}), we may calculate the absolute error $\bar{\delta}_f (x)$ similar to Eq.~(\ref{relative-error-static}) in Section~\ref{aberration-static}, i.e.
    \begin{eqnarray}
    \bar{\delta}_f (x) & \equiv & \left|\bar{f}(x)-\frac{R}{2 \gamma}\right| \nonumber \\
    & = & \frac{\left|f(x)-\frac{R}{2}\right|}{\gamma} = \frac{\delta_f (x)}{\gamma}.
    \label{absolut-error-moving}
    \end{eqnarray}
    Eq~(\ref{absolut-error-moving}) shows that the absolute error in the relativistic case differs from the static one by the factor $\frac{1}{\gamma}$. This means that when the mirror system is moving at a considerably high speed, the difference of $\bar{f}(x)$ from $\frac{R}{2 \gamma}$ is relatively small so that the position of the intersection point may be considered as a fixed point. Hence the spherical aberration of the relativistic mirror system may be reduced by increasing its speed. However, since the spherical aberration relation suffers the Lorentz contraction, the focal length of the mirror system moving with high speed will also be reduced according to Eq.~(\ref{focal-moving-conic}). As an illustration, Figure~\ref{aberration-moving-spherical} for the relativistic spherical concave mirror clearly indicates this effect. In the static case, as discussed in (\ref{paraxial}) of Section~\ref{aberration-static}, the paraxial region of a spherical mirror is $x < 0.304911$ with absolute error within $\delta_f \leq 0.025 R$. To attain the same absolute error within a larger range of $x$, we have to solve $\bar{\delta}_f (x) = 0.025 R$ for a specific value of $x$. For example, to attain the same absolute error of $0.025 R$ within the region $x < 0.5$, the mirror system must be moving with the speed $v > 0.946329 c$. In general, the minimal speed $v$ in which the absolute error is within the range $\bar{\delta}_f \le \delta_0$ and the paraxial region within $x < x_0$ for a given mirror shape with eccentricity $e$ is, from Eq.~(\ref{absolut-error-moving}),
    \begin{equation}
    v = c \sqrt{1-\left( \frac{\delta_0}{\delta_f (x)} \right)^2}.
    \label{moving-speed}
    \end{equation}
\end{enumerate}

\section{Conclusion and future work}
\label{conclusion}

We have discussed the properties of the relativistic curved mirrors and compared them with the properties of the static case. It is shown that the focal lengths of the relativistic mirrors are shorten according to the Lorentz contraction. Consequently, all focal lengths will tend to zero as speed of the mirror systems increases up to the speed of light (if it is possible to do so). From the Gaussian formula~(\ref{Taylor-4-moving}), we may conclude that the light rays are getting harder to reach the mirror system as the speed of the mirror system $v$ is approaching the speed of light, and hence the rays are getting harder to be reflected by the surface of the mirror ($\bar{S}_i \rightarrow 0$). The spherical aberrations are also affected by the movement of the mirror systems. The spherical aberration of the relativistic mirror system reduces as the speed of the system increases, but at the same time, the intersection position of the reflected ray on the principal axis reduces due to the Lorentz contraction. The parabolic mirror does not suffer any spherical aberration caused by any moving state, even though the associated focal length decreases with the increase of speed.

It will be interesting to analyze the more general situation where an observer is moving at some angle with respect to the principal axis. A preliminary construction of the associated OPL shows that there are some modifications of Eqs.~(\ref{OPL-moving-a}) and (\ref{OPL-moving-b}) in the form of
\begin{equation}
l_a (\bar{h}) = \bar{S}_0 \sqrt{\left[1 + \frac{v t_a \cos \alpha - \bar{l}(\bar{h})}{\bar{S}_0}\right]^2 + \left[\frac{\bar{h} - \bar{h}_0 + v t_a \sin \alpha}{\bar{S}_0}\right]^2}
\end{equation}
and
\begin{equation}
l_b (\bar{h}) = \bar{S}_i \sqrt{\left[1 - \frac{v t_b \cos \alpha + \bar{l}(\bar{h})}{\bar{S}_i}\right]^2 + \left[\frac{\bar{h}_i - \bar{h} + v t_b \sin \alpha}{\bar{S}_i}\right]^2},
\end{equation}
where $\alpha$ is the angle between the moving velocity and the principal axis, and $\bar{h}$ is the Lorentz contracted version of $h$ due to the speed $v \sin \alpha$. The detailed analysis of the situation will be discussed in a future paper.

It will also be interesting to investigate whether the same techniques used in this work can be applied to curved lenses. The work of Voronovich et.al. shows that Fermat's principle has proven to be valid for optical systems with moving optical media~\cite{Voronovich03,Godin04}. It is well known that every lens system suffers spherical aberration because of the change of light speed inside the lens' media. Logically, the Gaussian formula and the spherical aberration relation must change due to the change of OPL caused by the movement of the lens system. The work of Gjurchinovski et.al. shows that an isotropic optical medium may become optically anisotropic observed in a moving frame at a constant velocity~\cite{Gjurchinovski07}. Using the same techniques discussed here, we may discover how the speed of a lens system affects the associated focal length and spherical aberration, which can provide some new information about the behavior of relativistic lenses.

\section*{Acknowledgement}

The authors would like to thank Professor B. Suprapto Brotosiswojo, Dr. A. Rusli and Dr. Philips N. Gunawidjaja of the Department of Physics, Parahyangan Catholic University for the valuable comments, discussions and corrections during the early and final stages of this work. The authors would also like to thank the anonymous referees for the valuable comments and suggestions for the extension of the work in the future research. The work is partly supported by the research grant of the Directorate General of Higher Education - Ministry of National Education of The Republic of Indonesia.

\end{document}